\documentclass[conference]{IEEEtran}
\usepackage{blindtext, graphicx}
\usepackage{amsmath}
\usepackage{cite}
\ifCLASSINFOpdf
\else
\fi

\usepackage{xcolor}
\usepackage{multirow}
\usepackage[justification=centering]{caption}
\usepackage{booktabs,threeparttable}
\usepackage{footnote}
\usepackage{comment}
\usepackage{amssymb}% http://ctan.org/pkg/amssymb

\IEEEoverridecommandlockouts
\IEEEpubid{\makebox[\columnwidth]{978-1-5090-5499-2/17/\$31.00~ \copyright 2017 IEEE \hfill} \hspace{\columnsep}\makebox[\columnwidth]{ }}

\begin{document}
\title{Empirical comparison of three models for determining market clearing prices in Turkish day-ahead electricity market}

\author{\IEEEauthorblockN{G\"{o}khan Ceyhan}
\IEEEauthorblockA{Software R\&D Specialist\\
Energy Exchange Istanbul, Turkey\\
Email: gokhan.ceyhan@epias.com.tr}
\\
\IEEEauthorblockN{H. Bahadir Sahin}
\IEEEauthorblockA{Application Development Specialist\\
Energy Exchange Istanbul, Turkey\\
Email: bahadir.sahin@epias.com.tr}
\and
\IEEEauthorblockN{Nermin Elif Kurt}
\IEEEauthorblockA{Software R\&D Specialist\\
Energy Exchange Istanbul, Turkey\\
Email: nermin.kurt@epias.com.tr}
\\
\IEEEauthorblockN{K\"{u}r\c{s}ad Derinkuyu}
\IEEEauthorblockA{Dept. of Industrial Engineering\\
TOBB University of Economy and Technology, Turkey\\
Email: kderinkuyu@etu.edu.tr}}


%Telephone: (800) 555--1212, Fax: (888) 555--1212}


\maketitle

\begin{abstract}
%\boldmath
Bidders in day-ahead electricity markets want to sell/buy electricity when their bids generate positive surplus and not to take an action when the reverse holds. However, non-convexities in these markets cause conflicts between the actions that the bidders want to take and the actual market results. In this work, we investigate the non-convex market clearing problem of Turkish market operator and propose three different rule sets. The first rule set allows both rejection of bids with positive surplus and acceptance of bids with negative surplus. The second and the third sets only allow one of these conflicted cases. By using total surplus maximization as the objective, we formulate three models and statistically explore their performance with the real data taken from Turkish market operator. 

\end{abstract}


% \fi
\IEEEpeerreviewmaketitle

\section{Introduction}

Deregulation processes in the electricity industry have taken place in the last few decades. Starting in Chile 1982, it spread to many countries around the world including Turkey. In Turkey, the deregulation process started with the introduction of the first Electricity Market Law and the establishment of Electricity Market Regulatory Authority (EMRA) in 2001. The law was amended three times in 2012, 2013 and 2016. Currently, competition is in place at generation and wholesale layers.  In addition, retail market competition is expected to increase with gradually decreasing eligible consumer limits. 

The wholesale electricity market in Turkey has been operated by Energy Exchange Istanbul (EXIST) since its establishment in September, 2015. EXIST operates two electricity spot markets, day-ahead and intra-day. A daily double-sided blind auction is held in the day-ahead market under the principle of uniform pricing. On the other hand, intra-day market is operated under continuous trading mechanism. 

The day-ahead auction in Turkey is very similar to its counterparts in European markets. The participants of the auction can submit three types of bids: hourly, block and flexible. Hourly bids specify the quantity (production or consumption) offered to the market as a piecewise linear function of market clearing price. A block bid presents an indivisible amount of quantity for a single price and may be valid for a multiple consecutive periods as opposed to hourly bids. Lastly, flexible bids also represent an indivisible amount of quantity but only for a single period. As opposed to hourly and block bids, flexible bids are not submitted for a particular period and can be evaluated at any period by the market clearing algorithm. 

Given the set of submitted bids, the main function of the market operator is to determine the hourly market clearing prices and matching results for the bids (accepted quantity for an hourly bid, accept/reject decision for a block bid and the accepted period for a flexible bid if accepted). In the absence of indivisibility requirements of block and flexible bids, it is well-known that welfare maximizing solution clears the market and equilibrium prices exist. However, the indivisibility requirement transforms the market clearing problem into a non-convex problem such that equilibrium prices may not exist. That is, the marginal prices at the surplus maximizing solution may cause bidders to miss additional profits or even incur loss. In order to avoid such undesired market results, power exchanges may adopt additional constraints to the market clearing problem at the expense of suboptimal solutions in terms of market surplus.

The main challenge faced by EXIST is whether to integrate such rules to the market clearing problem. Without any additional constraints, the welfare maximizing solution may include undesired bid results such that a bidder who incurred a loss would be better off by not committing the matching quantity or a bidder who missed a profit would be better off by committing additional quantities. To remedy this problem at least partly, one option is to add a set of constraints to the market clearing problem to guarantee that no one incurs a loss. Similarly, another option could be to ensure that no one incurs a loss of profit at the expense of bids at loss. In this case, some compensation mechanism needs to be established to pay for the total loss of the bidders. Note that market loss and loss of profit cannot be prevented at the same time.

The widely accepted European solution is to prevent market loss. To do that, a common practice is to build a primal-dual formulation with complementarity constraints and relax some of them to guarantee the feasibility and no loss property. For example, the coupled markets of Europe are cleared by the algorithm named EUPHEMIA which satisfy this requirement \cite{euphemia}. In the literature, there are studies proposing formulations and solution methods for this problem, \cite{martin2014strict}, \cite{madani_vyve_2015} and \cite{madani2014minimizing}. In these studies, the authors present MILP or MIQP formulations (depends on the type of hourly bids: stepwise or piecewise) and branch-and-cut algorithms based on Bender's decomposition to solve those problems. 

In addition to the studies proposing solution methods for the current European pricing mechanism, there are some other studies discussing alternative pricing rules. For example, \cite{o2005efficient} suggests to price integral activities like start-ups as independent commodities complementary to the energy commodity. It means that when a supply block bid is accepted, the associated start-up price must be paid to the bidder in addition to the money received for the energy to be produced. \cite{van2011linear} examines and compares US and EU models in terms of optimality of market surplus, existence of Walrasian equilibrium, whether the bidders bid truthfully or the auctioneer encounters missing money problem. In addition, the author proposes a non-uniform pricing mechanism for the problem.

Recently, \cite{madani2016non} investigates applicability of non-uniform pricing proposed in \cite{van2011linear} for the European market in terms of economic, algorithmic and legal perspectives. The authors argue that uniform pricing may cause the rejection of some bids with potentially positive surplus and the increase in such occurrences at the last years challenge the confidence of the market participants on the optimality of the results. The model proposed allows the acceptance of bids at loss, but then those bidders are compensated (with \textit{uplifts} payments) by the ones having positive surplus. The non-uniform pricing algorithm is run on real instances of Belgium market and the results are compared with the ones under uniform pricing in terms of computation time, welfare and required uplifts. 

In the Turkish day-ahead market, the current pricing mechanism prevents the rejection of bids with potentially positive surplus. On the other hand, it allows to acceptance of bids at loss but those bids are then settled from their bid price instead of market clearing price. So, those bidders are fully compensated by the market operator. However, the resulting budget deficit of market operator is again financed by the market participants with a posterior procedure. 

Both Turkish and European models produce suboptimal solutions in terms of market total surplus. European practice sticks to uniform pricing despite the fact that some bids with very favorable prices compared to market clearing prices could be rejected. This practice may prevent to achieve market equilibrium. On the other hand, Turkish model deviates from uniform pricing but succeeds market equilibrium by compensating the bidders at loss.

In this study, we test three pricing mechanisms on past bid sets of EXIST and compare the results in terms of some performance measures interested by EXIST. Our ultimate aim is to help policy makers to choose the best practice for day-ahead market pricing. In particular, we would like to know whether the three pricing rules lead to significant differences on total market surplus and market clearing prices. We also try to estimate the cost of integrating rules such as ``no bid at loss" or ``no rejection of bids with favorable prices" into the market clearing algorithm in terms of the magnitude of deviation from optimal market surplus.

We contribute to the literature by presenting comprehensive empirical results for different pricing mechanisms on a large set of real instances. The computational results cover comparisons on performance measures like total surplus, number of bids accepted at loss or rejected with potential positive surplus and the price deviation of those bids from market clearing prices. Lastly, we propose a different formulation for the market clearing problem than the common European formulation.

In the next section, we present the necessary definitions to state the problem and give the notations. In Section \ref{models}, we give the mathematical models for the basic surplus maximization problem and two variants of it for European and Turkish markets. Afterwards, we present the results of our computational tests in Section \ref{cs} and discuss our findings in Section \ref{disc}. We conclude our paper in Section \ref{conc}.

\section{Definitions and Notations}

\begin {table}[h!]
\begin{tabular}{c l}
Indices and Sets & \\
\hline
$t,T $& Index and set of time periods. \\
$l, L$ & Index and set of hourly bid directions, \\
&where $l$ can be either supply (s) or demand (d), \\
&i.e., $L=\{s,d\}$ \\
$i,I_{lt} $& Index and set of hourly bid segment indices for $l$ and $t$. \\
$b,B$ & Index and set of block bids \\
\end{tabular}
\end{table}

\begin {table}[h!]
\begin{tabular}{c l}
Parameters &  \\
\hline
$P_{min},P_{max}$ & Minimum and maximum price limits. \\
$P^0_{ilt},P^1_{ilt}$ & Initial and final price for segment $i$, direction $l$ and period $t$, \\
$Q_{ilt}$ & Quantity of segment $i$ $\forall$ $i\in I_{lt}$, $l \in L$, $t\in T$.  \\
$P_b,Q_{bt}$ & Price and quantity of block bid $b$, $\forall$ $b\in B$, $\forall$ $t\in T $ \\
\end{tabular}
\end{table}

Hourly bids consist of a set of quantity price pairs where each pair shows the maximum/minimum price that a bidder is willing to buy/sell the corresponding quantity. The total quantity offered for a price level in a period is the sum of the corresponding quantities for each hourly bid. When buy and sell quantities are separated, the resulting aggregated quantity price pairs form a supply and a demand curve in the corresponding period. The former is a non-decreasing and the latter is a non-increasing function of quantity. Each area between the pairs forms a segment where $P^0_{ist} \leq P^1_{ist}$ for all $i\in I_{st}$, $t\in T$ and $P^1_{idt} \leq P^0_{idt}$ for all $i\in I_{dt}$, $t\in T$. Quantity of segment $i$ is the difference of quantities of consecutive pairs where $Q_{ist} \leq 0$ for all $i\in I_{st}$, $t\in T$ and $Q_{idt} \geq 0$ for all $i\in I_{dt}$, $t\in T$.

\begin {table}[h!]
\begin{tabular}{c l}
Decision Variables &\\
\hline
$x_{ilt}$ & Accepted fraction of segment $i$, $\forall$ $i\in I_{lt}$, $l \in L$, $t\in T$. \\
$y_b$ & 1 if block bid $b$ is accepted, 0 otherwise, $\forall$ $b\in B $.\\
$p_t$ & Market clearing price at period $t$. \\
$w_{ilt}$ & 1 if segment l at period t is fully accepted, 0 otherwise.\\
\end{tabular}
\end{table}

Given market clearing price vector $\textbf{p}=[p_1, p_2, ...]$, a block bid $b \in B$ is \textit{in-the-money} if $\sum\limits_{t \in T} Q_{bt}(P_b - p_t)>0$, it is \textit{out-of-the-money} if $\sum\limits_{t \in T} Q_{bt}(P_b - p_t)<0$ and it is \textit{at-the-money}, otherwise. If an in-the-money block bid is rejected, then it is called \textit{paradoxically-rejected-bid (PRB)}. On the other hand, if an out-of-the-money block bid is accepted, then it is called \textit{paradoxically-accepted-bid (PAB)}. Let $B_0$ and $B_1$ be the set of PRBs and PABs, respectively. Then, we define \textit{total loss (TL)} and \textit{total loss of profit (TLP)} as follows:  

\begin{equation}
     TL=-\sum\limits_{b\in B_1} \sum\limits_{t \in T} Q_{bt}(P_b - p_t)
\end{equation}

\begin{equation}
     TLP=\sum\limits_{b\in B_0} \sum\limits_{t \in T} Q_{bt}(P_b - p_t)
\end{equation}

Similarly, we also need to define maximum loss (MUL) and maximum loss of profit (MULP) per unit energy in order to assess the fairness of auction result among the bidders.

\begin{equation}
     MUL=\max\limits_{b\in B_1} \big| \sum\limits_{t \in T} (P_b - p_t)\big|
\end{equation}

\begin{equation}
     MULP=\max\limits_{b\in B_0} \big| \sum\limits_{t \in T} (P_b - p_t) \big|
\end{equation}

\section{Market Clearing Models}
\label{models}

We only consider hourly and block bids where the latter signal all-or-nothing requirements of the participants, therefore, create non-convexities in the system. We exclude flexible and linked block bids for the sake of simplicity of the presentation.

The type of the hourly bids determines the type of the objective function, hence the type of the model. If the quantity offered by an hourly bid is a stepwise function of price, then the model is a mixed integer linear program (MILP). On the other hand, if bid quantity is a piecewise linear function of the price, then the objective function is quadratic, and the problem becomes mixed integer quadratic program (MIQP). In this study, we assume that all hourly bids are piecewise linear functions so that the problem is MIQP. 

The well-known non-convex surplus maximization model (M1) is constrained by only the clearing constraints (supply-demand balance). The non-convexity of (M1) is due to all-or-nothing nature of the block bids. Thus, there may occur both PRBs and PABs at the optimal solution of (M1) failing to achieve Walrasian equilibrium. Adding constraints to (M1) which eliminate both PAB and PRBs is not possible since the resulting problem may be infeasible under block bid integrality constraints. Therefore, power exchanges assent one of these requirements. 

In the following part, we first present (M1). Then, we give the two variant models in which either PRBs or PABs are allowed.

\subsection{Case 1: Allowing both paradoxically accepted and rejected block bids}
In this section, we formulate model (M1) where there is only supply and demand balance constraint in addition to the bounds on the segment variables.

(M1)
\begin{equation*}
\begin{split}
    \max \sum_{t\in T} \Big ( \sum_{l\in L} \sum_{i\in I_{lt}} Q_{ilt}P_{ilt}^0 x_{ilt} &+ Q_{ilt}(P_{ilt}^1-P_{ilt}^0)\frac{x_{ilt}^2}{2} \\
    &+ \sum_{b\in B} Q_{bt} P_b y_b \Big )\\
\end{split}
\end{equation*}

\begin{align}
     &\sum_{l\in L} \sum_{i \in I_{lt}} Q_{ilt}x_{ilt} + \sum_{b\in B} Q_{bt} y_b = 0, \quad \forall t\in T \label{balanceC}\\
     & x_{ilt} \leq 1, \quad \forall i \in I_{lt},  t\in T , l\in L \label{segmentRatioUpperBoundC} \\
     & x_{ilt} \geq 0, \quad\forall i \in I_{lt},  t\in T , l\in L , \quad  y_b \in \{0,1\} \quad \forall b \in B \label{variables1C}
\end{align}

Constraint \eqref{balanceC} defines supply-demand balance for each period, \eqref{segmentRatioUpperBoundC} ensures that accepted fraction of a segment does not exceed 1 and \eqref{variables1C} includes variable non-negativity and integrality constraints.

\subsection{Case 2: Allowing only paradoxically rejected block bids}
One alternative of a market operator is to allow PRBs and eliminate PABs. In this setting, the market operator does not have a missing money problem and all energy is cleared from the same price. This is the widely accepted European solution. We refer to this model as (M2).

(M2) is formulated by adding constraints \eqref{forceNotPABC}-\eqref{variables2C} to model (M1).

\begin{equation}
 \label{forceNotPABC}
     M_b (y_b-1)\leq \sum_{t\in T} Q_{bt} (P_b-p_t), \\   
\end{equation}
where $M_b= \sum_{t\in T} Q_{bt} (P_{max}-P_{min})$. When bid $b$ is \textit{out-of-the-money}, i.e., $\sum_{t\in T}Q_{bt} (P_b-p_t) < 0$, bid $b$ is rejected, but when it is \textit{in-the-money}, $b$ may be accepted or rejected. Hence, \eqref{forceNotPABC} forces \textit{out-of-the-money} bids to reject.

In this model and the following one, we need to integrate the market clearing price variables into the problem. This, in turn, necessiates to explicitly model hourly bid equilibrium constraints.The following constraint relates $p_t$ and hourly bid equilibrium quantities.

\begin{equation}
    \label{mcpC}
    \begin{split}
         & p_t = P_{min} + \sum_{i\in I_{st}} (P_{ist}^1-P_{ist}^0){x_{ist}}, \quad \forall t\in T \\
         & p_t = P_{max} + \sum_{i \in I_{dt}} (P_{idt}^1-P_{idt}^0){x_{idt}}, \quad \forall t\in T \\
     \end{split}
\end{equation}

Constraint (10) guarantees that an hourly bid segment cannot be accepted unless the preceding one is accepted at full.

\begin{equation}
    \label{continuityC}
    \begin{split}
         & w_{ilt} \leq x_{ilt} \leq w_{(i-1)lt},\quad \forall i\in I_{lt}, t \in T,  l\in L  \\
         & w_{0lt} = 1, \quad w_{|I_{lt}|lt} = 0 \quad \forall t \in T,  l\in L  \\ \\    
     \end{split}
\end{equation}

\begin{equation}
    \label{variables2C}
   w_{ilt} \in \{0,1\}, \quad\forall i \in I,  t\in T , l\in L
\end{equation}

\subsection{Case 3: Allowing only paradoxically accepted block bids }
Another alternative is to prevent PRBs and allow PABs. This is the rule in the current Turkish market. The PABs are settled from the bid price so that their loss is compensated. This creates the missing money problem, which is called \textit{side payment} in Turkish market context, but supports equilibrium in the market. Both bidders having PAB or PRB cannot be better off by changing the commitments determined by the market operator. We refer this model as model (M3).

(M3) is formulated by adding constraints \eqref{mcpC}-\eqref{forceNotPRBC} to model (M1).

\begin{equation}
    \label{forceNotPRBC}
   \sum_{t\in T} Q_{bt} (P_b-p_t) \leq M_b y_b,
\end{equation}

where $M_b= \sum_{t\in T} Q_{bt} (P_{max}-P_{min})$. While $b$ has positive surplus then $y_b$ must be 1, therefore, \eqref{forceNotPRBC} eliminates PRBs. Yet, when $b$ has negative surplus then $y_b$ may still take value 1, which may cause PAB. 

\section{Computational Study}
\label{cs}

In this section, we first summarize the market data used in the experiments. Then, we briefly explain the performance measures used to compare three pricing models stated in Section \ref{models}. Afterwards, we specify the scope of our computational tests and present the results in a comparative manner.

\subsection{Data}
In our computational experiments, we use real bid data sets of Turkish day-ahead market. The instances cover the period from the very beginning of the market in 2012 up to midst of 2016. The total number of instances used in the experiments reach up to around 1600. The bid data set includes private pricing information of market players and therefore is not publicized. In the following two tables, we present the main characteristics of the Turkish day-ahead market bid sets in terms of daily average bid counts and volumes. 

Table \ref{bidcount} reveals that the daily average number of bids in the auction does not show a clear increasing pattern over the years. On the contrary, the number of demand block bids decreased sharply from 2014 to 2015. A majority of block bids are offered to sell energy to the market. 

Apart from the bid counts, we see in Table \ref{bidvol} that total supply volume is more than total demand volume offered to the market in the last years. In addition, the share of hourly bids in supply volume decreasing slightly as opposed to the increasing share of hourly bid volumes in demand.

\begin{savenotes}
\begin{table}[ht!] 
\centering
\begin{threeparttable}[b]
\caption{The daily average number of bids in Turkish DAM with respect to bid types} 
\label{bidcount}
\centering
\begin{tabular*}{0.48\textwidth} {@{\extracolsep{\fill}}  *{1}{r}   *{1}{r}  *{1}{r} *{1}{r}  }
\toprule

\multicolumn{1}{r}{Year}	& \multicolumn{1}{r}{Hourly Bids} & \multicolumn{1}{r}{Supply Block Bids} & \multicolumn{1}{r}{Demand Block Bids} \\
\midrule
2012	&	7,323	&	87  &   46	\\
2013	&	8,808	&	107  &   42	\\
2014	&	10,064	&	99  &   38	\\
2015	&	9,815	&	117  &   15	\\
\bottomrule
\end{tabular*}

\end{threeparttable}
\end{table} 
\end{savenotes}

\begin{savenotes}
\begin{table}[ht!] 
\centering
\begin{threeparttable}[b]
\caption{The daily average volume of supply and demand bids in Turkish DAM with respect to bid types} 
\label{bidvol}
\centering
\begin{tabular*}{0.48\textwidth} {@{\extracolsep{\fill}}  *{1}{r}   *{2}{r}  *{2}{r} *{1}{r}  }
\toprule
&    \multicolumn{2}{ c }{Supply (MWh)} & \multicolumn{2}{ c }{Demand (MWh)} \\
\cmidrule(lr){2-3} \cmidrule(l){4-5}
\multicolumn{1}{r}{Year}	& \multicolumn{1}{r}{Hourly Bids} & \multicolumn{1}{r}{Block Bids}	& \multicolumn{1}{r}{Hourly Bids} & \multicolumn{1}{r}{Block Bids} \\
\midrule
2012	&	259,873 (80\%) &    64,680 (20\%)  &   249,517 (64\%) &   139,439 (36\%)	\\
2013    &   305,585 (80\%) &    86,638 (20\%)  &   268,786 (71\%) &   112,213 (29\%) \\
2014    &   365,889 (78\%) &   104,438 (22\%)  &   312,593 (72\%) &   122,798 (28\%) \\
2015    &   384,084 (72\%) &   150,849 (28\%)  &   364,695 (92\%) &    30,535 (8\%)  \\
\bottomrule
\end{tabular*}
\begin{tablenotes}{\footnotesize \item [*] Values in parentheses show the ratio of corresponding volumes to the total supply or demand volumes.}
\end{tablenotes}
\end{threeparttable}
\end{table} 
\end{savenotes}

\subsection{Performance Measures}
In a convex market clearing problem, total surplus maximizing solution is the best economic solution. It clears the market at equilibrium prices and do not lead to market loss or loss of profit. If this is the case, the auctioneer does not need any other criteria to evaluate the auction results. However, in non-convex markets like today's day-ahead electricity markets, the market operator needs to take into account multiple and most of the times conflicting objectives. In EXIST, the scores of auction result in the following performance measures are monitored on a daily basis:

\begin{savenotes}
\begin{table}[ht!] 
\centering
\begin{threeparttable}[b]
\caption{Performance measures} 
\label{measures}
\centering
\begin{tabular*}{0.48\textwidth} {@{\extracolsep{\fill}} *{1}{l} *{1}{c}  }
\toprule

%\multicolumn{1}{l}{Measure} & \multicolumn{1}{c}{Abbreviation} \\
%\midrule
Total surplus & \textit{TS}  \\
Daily average of market clearing prices & \textit{MCP} \\
Number of PABs & \textit{\#PAB} \\
Number of PRBs & \textit{\#PRB} \\
Total loss (incurred by PAB bids) & \textit{TL} \\
Total loss of profit (missed profit by PRBs) & \textit{TLP} \\
Maximum loss per unit energy & \textit{MUL} \\
Maximum loss of profit per unit energy & \textit{MULP} \\
\bottomrule
\end{tabular*}

\end{threeparttable}
\end{table} 
\end{savenotes}

The smaller is the better for all criteria except the first and second one. For EXIST, occurrences of PABs or PRBs are equally undesirable. Hence, we do not differentiate between number of PABs or PRBs and associated loss or loss of profit values. That is, we consider the sum of these values when interpreting the results associated with model (M1). In other models, at least one of those values are zero anyway. 

\subsection{Tests}

We apply three pricing models given in Section \ref{models} to each instance in our sample. We call the MIQP solver of IBM ILOG CPLEX 12.6.3 to solve the associated MIQP problems. We set the absolute gap tolerance to 100 Turkish liras (1 Turkish lira is around 0.26 Euros at the time of this writing.), time limit to 120 seconds and left other solver parameters at their default values. We exclude an instance from the comparisons if any of three problems could not be solved to the optimality under the given time limit. In addition, we exclude the cases if price caps became binding in the optimal solution of at least one of the three cases. Thus, we try to measure the pure effect of the pricing rules by eliminating possible distortions by the sub-optimality of the solutions. The number of instances where we were able to find the optimal solutions for all three cases is 1538 out of 1631.

At the end of each run, we store the values of performance measures at the optimal solution found. We did not take into account the existence of alternative optima and no posterior procedure was implemented to pick a specific one of them. For each performance measure $k$, we build the following hypotheses separately:

\begin{equation}
    H_{12}^k:\mu(M1_k)-\mu(M2_k)=0   
\end{equation}
\begin{equation}
     H_{13}^k:\mu(M1_k)-\mu(M3_k)=0
\end{equation}
\begin{equation}
     H_{23}^k:\mu(M2_k)-\mu(M3_k)=0
\end{equation}

We test the hypotheses by using \textit{paired t-test} with Type-1 error set to 0.05, i.e. $\alpha = 0.05$. We report the associated $p-$value statistics and $(1-\alpha)$ confidence intervals on the population means. 

\subsection{Results}
In Table \ref{results1}, we report the results of the hypotheses as well as the sample mean differences, $\bar{X}$. Last two columns show the 95\% confidence limits where LCL and UCL corresponds to lower and upper confidence limit, respectively. $p-$values smaller than $\alpha$ show that the corresponding hypothesis is rejected whereas larger values imply that we fail to reject the given hypothesis. Following figures further expose the dispersion of differences among model pairs for each performance criterion. In the given box-plots, horizontal borders of the boxes correspond to first, second and third quartile,  respectively. The lowest and highest lines, called whiskers, are 1.5 times the inter-quartile range away from the boxes. The data points outside the whiskers are called outliers which are not shown on some of the figures for the sake of display quality. 

\begin{savenotes}
\begin{table}[ht!] 
\centering
\begin{threeparttable}[b]
\caption{Test statistics reported for the hypotheses build on each performance measure and pair of models} 
\label{results1}
\centering
\begin{tabular*}{0.48\textwidth} {@{\extracolsep{\fill}}  *{1}{l} *{1}{c} *{4}{r} }
\toprule
\multicolumn{1}{l}{Criterion ($k$)}	& \multicolumn{1}{c}{Hypothesis} & \multicolumn{1}{r}{$\bar{X}$}	& \multicolumn{1}{r}{$p-$ value} & \multicolumn{1}{r}{LCL} & \multicolumn{1}{r}{UCL} \\
\midrule
\multirow{3}{*}{TS} 	&	$H^k_{12}$ &    2,375  &   0 &   2,111 & 2,639 \\
&	$H^k_{13}$  &    2,256  &   0 &   2,000 & 2,512 \\
&	$H^k_{23}$  &    -119  &   0.52 &   -485 & 248 \\
\midrule
\multirow{3}{*}{MCP} 	&	$H^k_{12}$ &    -1.14  &   0 &   -1.26 & -1.02 \\
&	$H^k_{13}$  &    0.93  &   0 &   0.82 & 1.04 \\
&	$H^k_{23}$  &    2.07  &   0 &   1.9 & 2.23 \\
\midrule
\multirow{3}{*}{\#PAB + \#PRB} 	&	$H^k_{12}$ &    -0.14  &   0 &   -0.20 & -0.08 \\
&	$H^k_{13}$  &    -0.16  &   0 &   -0.23 & -0.09 \\
&	$H^k_{23}$  &    -0.02  &   0.61 &   -0.11 & 0.07 \\
\midrule
\multirow{3}{*}{TL + TLP} 	&	$H^k_{12}$  &   -108,774  &   0 & -122,425   & -95,122 \\
&	$H^k_{13}$ &    62,997  &   0 &   53,775 & 72,218 \\
&   $H^k_{23}$  &   171,770 &   0   &   155,446    &    188,096\\   
\midrule
\multirow{3}{*}{Max. price diff.*} 	&	$H^k_{12}$  &   -2.85  &   0 & -3.25   & -2.45 \\
&	$H^k_{13}$ &    -3.08  &   0 &  -3.51 & -2.64 \\
&   $H^k_{23}$  &   -0.23 &   0.44   &   -0.80    &    0.35\\   

\bottomrule
\end{tabular*}
\begin{tablenotes}{\footnotesize \item [*] Max. price diff. = max$\left\{MUL,MULP\right\}$}
\end{tablenotes}
\end{threeparttable}
\end{table} 
\end{savenotes} 

\begin{figure}[ht!]
\begin{center}
\includegraphics[width=0.48\textwidth]{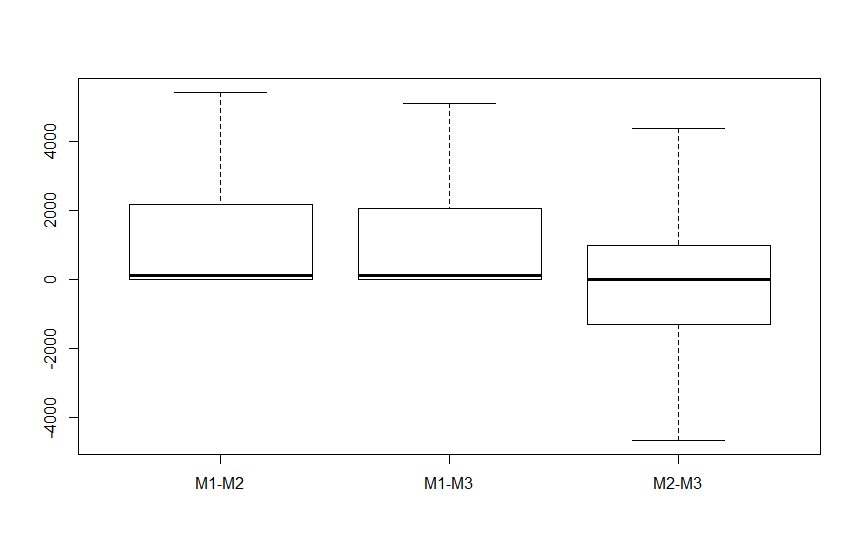}
\caption{Box plots for the total surplus differences among the model pairs} 
\label{f1}
\end{center}
\end{figure}

First of all, Table \ref{results1} shows that there is no significant difference in total surplus between model (M2) and (M3). In addition, it seems that pricing with model (M2) or (M3) costs minor surplus losses which are expected to be around 2,000 Turkish liras per day on the average. Figure \ref{f1} shows the spread of total surplus differences among the model pairs. In all cases, medians are very close to zero and (M2)-(M3) differences seem to distribute around zero symmetrically. Although not shown on the figure, we would like to note that extreme outliers at both directions may reach up to 50,000 and -50,000 Turkish liras.

\begin{figure}[ht!]
\begin{center}
\includegraphics[width=0.48\textwidth]{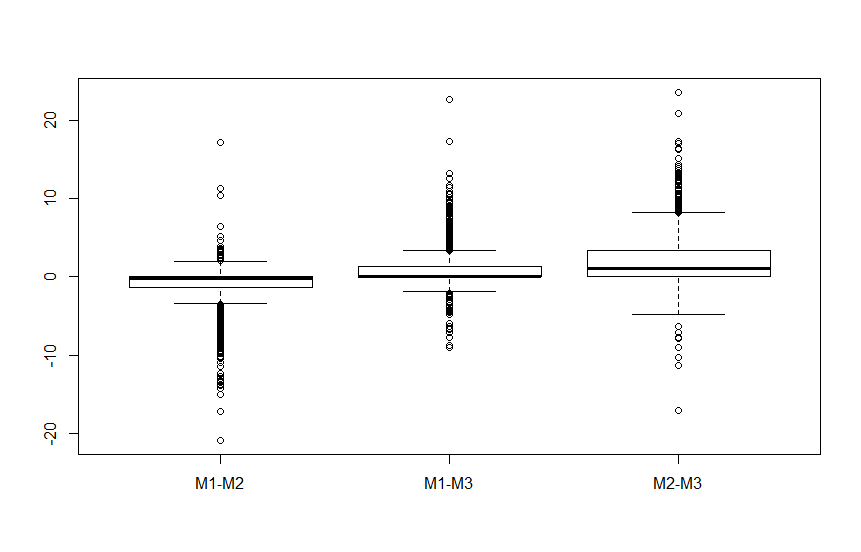}
\caption{Box plots for the daily average MCP differences among the model pairs} 
\label{f2}
\end{center}
\end{figure}

In terms of market clearing prices, we can argue that model (M2) increases the daily average around 2 Turkish liras compared to model (M3) on the average. This can be attributed to the large proportion of supply block bid volume in the total block bid volume (around 83\% in 2015). Since model (M3) is expected to accept more block bids compared to (M2) and most of them are supply bids, market clearing prices tend to decrease with model (M3). As Figure \ref{f2} shows, there are more outliers on the upper side of the last box plot reaching up to 20 Turkish liras. 

Our computational tests revealed that at 20\% of all cases, optimal solution of (M1) did not lead to any PAB or PRB. That is, three models share the same optimal solution at the 20\% of the cases. In addition, in 42\% of cases there were no PABs and in 45\% of cases there were no PRBs at the optimal solution of (M1). Although t-test statistics show that (M1) generates less paradoxical bids on the average compared to other models, the difference is very small. In addition, no significant difference has been found between average number of PRBs in (M2) and average number of PABs in (M3).

%\begin{figure}[ht!]
%\begin{center}
%\includegraphics[width=0.48\textwidth]{PB}
%\caption{Box plots for the paradoxical bid count (\#PAB + \#PRB) differences among the model pairs} 
%\label{f3}
%\end{center}
%\end{figure}

We see in Table \ref{results1} that the total loss of profit missed by PRBs dominates the actual loss incurred by PABs. (M3) significantly reduces the sum of those two amounts on the average whereas (M2) increases it.

%\begin{figure}[t]
%\begin{center}
%\includegraphics[width=0.48\textwidth]{TL}
%\caption{Box plots for the TL+TLP differences among the model pairs} 
%\label{f4}
%\end{center}
%\end{figure}

\begin{figure}[ht!]
\begin{center}
\includegraphics[width=0.48\textwidth]{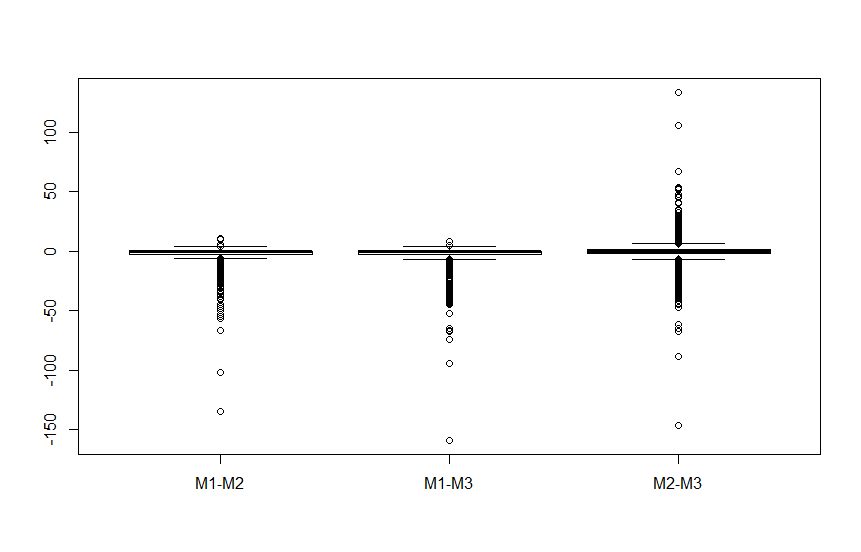}
\caption{Box plots for the maximum paradoxical bid price differences among the model pairs} 
\label{f5}
\end{center}
\end{figure}

A striking result comes at the price differences between paradoxical bid prices and the average market clearing prices of the periods where those bids are active. Test results show that both (M2) and (M3) increases MUL or MULP on the average compared to (M1). Actually, the differences can become quite large. Figure \ref{f5} shows that both MULP and MUL in models (M2) and (M3) may extend up to 150 Turkish liras more compared to (M1). We should note that maximum average market clearing price in our sample is 340 Turkish liras. In case of (M1) pricing, the maximum of MUL and MULP in the sample is at most 30 Turkish liras. This observation is also noted by \cite{van2011linear} to criticize the fairness of European model and mainly attributed to the existence of block bids with small volumes together with other block bids very large in the volume.

\section{Discussion}
\label{disc}
Our computational results show that integrating market pricing rules to prevent bidders from loss or loss of profit do not lead to substantial surplus loss on the average for the historical bid set of Turkish day-ahead market. However, for the market instances with higher percentages of block bid volumes, the differences can be expected to be larger. Similarly, average market clearing prices do not change too much on the average with the experimented pricing models although outlier price differences as much as 20 Turkish liras exist. But, the real cost of those additional pricing rules shows up as rejection of too deep in-the-money bids or acceptance of too far out-of-the-money bids. 

In \cite{van2011linear}, the author compares US and EU models in terms of welfare maximization, Walrasian equilibrium, truth-telling, fairness among bidders and market operator's missing money. Convex auctions are welfare maximizing, achieves Walrasian equilibrium, encourages bidders to bid truthfully, generates fair results and the amount of money received and paid by the market operator is always equal (no missing money). In addition, achieving the optimal solution is rather easy.

For the nonconvex auctions in the European markets including Turkey, all of the desired properties mentioned above is hard to satisfy if not impossible at all. In Table \ref{modelCompare}, we compare the pricing models presented here in terms of those characteristics. Pricing by (M2) and (M3) are not welfare maximizing due to the additional constraints in the models. (M3) assures Walrasian equilibrium since the in-the-money bids are not rejected and accepted out-of-the-money bids are compensated. However, it may not give strong incentives to bidders to bid truthfully since it is possible to have some bids settled from more favorable prices than the market clearing prices. If we define fairness as the magnitude of maximum loss or missed profit per unit of energy, we would like it to be as small as possible. In this regard, we can claim that pricing with (M1) results in more fair results compared to (M2) or (M3).

\begin{savenotes}
\begin{table}[ht!] 
\centering
\begin{threeparttable}[b]
\caption{The comparison of three pricing models} 
\label{modelCompare}
\centering
\begin{tabular*}{0.40\textwidth} {@{\extracolsep{\fill}} *{1}{l} *{3}{c}  }
\toprule

\multicolumn{1}{l}{} & \multicolumn{1}{c}{M1}	& \multicolumn{1}{c}{M2} & \multicolumn{1}{c}{M3} \\
\midrule
Welfare Maximizing & \checkmark &  &  \\
Walresian Equilibrium & & & \checkmark \\
Truth-telling & & \checkmark & \\
Fair & \checkmark & & \\
No missing money & & \checkmark & \\
\bottomrule
\end{tabular*}

\end{threeparttable}
\end{table} 
\end{savenotes}

\section{Conclusion}
\label{conc}
We have presented three different models for Turkish day-ahead electricity market by considering only hourly and block bids which constitute almost all of the Turkish day-ahead market volume. By using real data taken from the market operator, the models are solved and their performances are compared in terms of total surplus, average market clearing prices, number of PABs and PRBs, total loss and loss of profit and maximum absolute difference between matching and market clearing prices. Our statistical experiments revealed insightful results for the Turkish market operator on the efficiency of the current and alternative market designs. According to paired t-test results, we fail to reject the hypothesis that there is a difference between (M2) and (M3) in terms of average total surplus, but the average market clearing prices differ around 2 Turkish liras. Moreover, total loss and loss of profit changes significantly among (M1), (M2) or (M3). We also would like to know how the results change when those characteristics change. For this purpose, we will conduct similar experiments on specially generated synthetic data. Furthermore, as EXIST plans to integrate new bid types to the market, we plan to extend the models and statistical tests correspondingly.

% or


%\appendices
%\section{Proof of ...}

%\section*{Acknowledgment}


\ifCLASSOPTIONcaptionsoff
  \newpage
\fi



\bibliographystyle{IEEEtran}
\bibliography{EEM}
%\begin{thebibliography}{1}

%\end{thebibliography}

%\begin{biography}[{\includegraphics[width=1in,height=1.25in,clip,keepaspectratio]{mshell}}]{Michael Shell}


%\vfill


\end{document}